\documentclass[letterpaper, 10 pt, conference]{ieeeconf}  

\IEEEoverridecommandlockouts                              

\usepackage{graphics}
\usepackage{graphicx} 
\usepackage{amsmath} 
\usepackage{amssymb}  
\usepackage{caption}

\title{\LARGE \bf
Home Automation Using SSVEP \& Eye-Blink Detection Based Brain-Computer Interface
}

\author{Kratarth Goel$^{1}$, Raunaq Vohra$^{2}$ , Anant Kamath$^{3}$ and Veeky Baths$^{4}$
\thanks{$^{1}$Kratarth Goel is with the Department of Computer Science, BITS Pilani KK Birla Goa Campus, Goa, India
        {\tt\small kratarthgoel@gmail.com}}%
\thanks{$^{2}$Raunaq Vohra is with the Department of Mathematics, BITS Pilani KK Birla Goa Campus, Goa, India
        {\tt\small ronvohra@gmail.com}}%
\thanks{$^{3}$Anant Kamath is with the Department of Electrical and Electronics Engineering, BITS Pilani KK Birla Goa Campus, Goa, India
		{\tt\small kamathanant@gmail.com}}%
\thanks{$^{4}$Veeky Baths is with the Department of Biology, BITS Pilani KK Birla Goa Campus, Goa, India
        {\tt\small veeky@goa.bits-pilani.ac.in}}%
}

\begin{document}

\maketitle
\thispagestyle{empty}
\pagestyle{empty}

\begin{abstract}
In this paper, we present a novel brain computer interface based home automation system using two responses - Steady State Visually Evoked Potential (SSVEP) and the eye-blink artifact, which is augmented by a Bluetooth based indoor localization system, to greatly increase the number of controllable devices. The hardware implementation of this system to control a table lamp and table fan using brain signals has also been discussed and state-of-the-art results have been achieved.
\end{abstract}
\smallskip

\noindent \textbf{\emph{Keywords -}}{\emph{brain-computer interface; assistive technologies; intelligent environments; ubiquitous computing}}

\section{Introduction}

This paper aims at implementing an efficient and effective brain-computer interface (BCI) based system with bluetooth based indoor user localization that enables users to control various home appliances without actual physical interaction. For a wide range of people, from those who are almost completely paralyzed to completely healthy people, we provide a non-graphical user interface (GUI) hardware based implementation that frees them from the hassle of pre-training or intermittently staring at a screen.  Our implementation consists of a combination of two responses - Steady-State Visually Evoked Potential (SSVEP) and Eye-Blink artifacts. 

\section{The Algorithm}

In this section, we discuss the algorithms for the detection of both the SSVEP response and the eye-blink detection artifact. 

\subsection{SSVEP Detection}

The useful frequency range to induce an adequate SSVEP response is limited to an interval of 6-24Hz, and the inter-stimulus gap has to be greater than or equal to 0.2Hz. But this frequency interval is enough to assign an independent frequency to a large number of devices. When a user focuses on a light source flickering at a predetermined constant frequency within the aforementioned interval, the evoked EEG signal can be used to determine the device associated with that frequency. The successful selection of this device can be used to trigger the desired action. The EEG data sample of 2 seconds at the sampling rate of 512Hz for SSVEP detection is obtained by making $O_{2}$ the signal electrode, $P_{3}$ the ground electrode and $F_{4}$ the reference electrode.
We define a set of target frequencies $F$, given by,

\begin{equation}
F = \{f_1 , f_2, .. , f_m\}
\end{equation}
Each $f_k$ is the target frequency for controlling the $k\textsuperscript{th}$ device out of a total of $m$ devices. After sampling for 2 seconds, autocorrelation is applied on the raw EEG data to reduce the noise. Fast Fourier Transform (FFT) is then applied to the results obtained after the autocorrelation step. We then calculate $A_k$ as the sum of the power amplitudes $|P|$ around the target frequencies $f_k$ and its second harmonic $2f_k$, $k \in [1,m]$. For the purposes of this experiment, $m = 2$ (one for the table lamps and one for the table fans). We calculate the sum of power amplitudes around the two chosen target frequencies of 6Hz for lamps and 8.2Hz for fans respectively. $A_k$ is given by,

\begin{equation}
A_k = \sum\nolimits_{f_k-0.05}^{f_k+0.05}  |P| + \sum\nolimits_{2f_k-0.05}^{2f_k+0.05}  |P|
\end{equation}

We consider the second harmonic because they are known to elicit a response equal to or stronger than the fundamental frequency \cite{A19}. In order to detect SSVEP responses, we use \emph{adaptive thresholding} on the values of $A_k$. The adaptive threshold $\tau$ is given by,

\begin{equation}
\tau = c\left\{\frac{1}{m}\sum_{k=1}^{{m}}A_k\right\}
\end{equation}

where $c \in \mathbb{R}$, can be taken as a parameter to adjust the threshold sensitivity. For the purposes of this experiment, we take the value of $c$ as 2. It must be noted here that this threshold parameter $\tau$ is calculated over a 4 second sample, rather than a standard 2 second sample used for the calculation of $A_k$. This is done so that the peaks detected in this sample are not localized. The frequency $f_k$ corresponding to the maximum value of $A_k$ ($A_{max}$) obtained for each sample  is selected as the target frequency, under the condition that the value of $A_{max}$ is greater than the threshold $\tau$ for the sample. The sensitivity parameter $c$ can therefore be adjusted accordingly. 

\subsection{Eye-Blink Artifact Detection}

After SSVEP is detected, which indicates that device selection has occurred, a 4 second window is provided in which the user blinks thrice in order to confirm his device selection. If such a sequence is detected, then the state of the selected device is toggled. This adds an extra layer of error resistance to the system, since the user has the option to not blink thrice in case SSVEP has been erroneously evoked. In order to detect voluntary eye blinking in the raw EEG data, we first apply a Butterworth 4\textsuperscript{th} order digital band pass filter in the range of 1-10Hz on the signal, which is obtained from the $F_{p2}$ electrode position. Then, using the same argument as for SSVEP, we apply adaptive thresholding to detect signal peaks. This threshold $\sigma$ is given by,

\begin{equation}
 \sigma = c'\left\{\frac{1}{n}\sum_{j=1}^{{n}}S_j\right\}
\end{equation}
where $S_j$ is the input signal potential at the $j\textsuperscript{th}$ time instant, $n$ is the number of samples (1024 samples in a 2 second window at a sampling rate of 512Hz) and $c' \in \mathbb{R}$ can be thought of as a sensitivity parameter for $\sigma$. For the purposes of this experiment, the value of $c'$ has been taken as 5. After identifying the peaks above the threshold $\sigma$, we then check whether they have peak width greater than 200ms. This is done to distinguish between a voluntary eye blink and random noise which could include involuntary eye blinking. 

\subsection{User localization}

The limited set of stimulating frequencies that can be used for SSVEP detection limits the number of devices that can be controlled using this system. One technique to increase the number of controllable devices is to use indoor user localization. Localization allows reuse of frequencies among multiple areas. Let us assume, for instance, that the number of detectable SSVEP frequencies is $N$. Without localization, mapping one frequency to one device allows control of $N$ devices. If the coarse localization system can resolve the user's location into $n$ discrete sub-areas withing the indoor environment, we can effectively increase the total number of controllable devices to $n$ x $N$. Coarse room level indoor user localization is performed by comparing the Received Signal Strength Indication (RSSI) reported by bluetooth beacons placed in each room, each of which tries to establish an L2CAP layer connection with the bluetooth radio on the neuroheadset, as described in \cite{A22}. This particular technique has the advantage of not requiring bluetooth pairing between the beacons and the radio on the headset. The beacons themselves are networked with the main processing computer, and relay these RSSI values to the computer. The system localizes the user to be in the room where the beacon corresponding to that room reports the highest RSSI value. 

\section{The Experiment}

In this experiment, we use our algorithm for controlling two devices - a table fan and a table lamp - in each room. Thus there are a total of 2 fans and 2 lamps being controlled in this experiment. Also, we use a cluster of six 5mm LEDs  for generating the stimulus used to control each device and HC-04 . We limit ourselves to two frequencies in this experiment, 6Hz and 8.2Hz. The cluster of LEDs for the fans in each room is made to flicker at a frequency of 8.2Hz and the LEDs for both the lamps flicker at 6Hz. The LED clusters are controlled usbluetooth modules for localization. In order to select a device, the subject needs to look at the respective LED cluster so that the EEG data corresponding to that period in time may show SSVEP response. The system also provides feedback to the user by pausing the flicker action for the duration of selection and causing the LEDs to stop flickering. If this feedback appears at time $t$ seconds, the subject can confirm his device selection by blinking his eyes 3 times by time ($t + 4$) seconds. This is done to ensure that the SSVEP based device selection was not done accidentally by either the subject or the algorithm, thus making the system more robust. The localization system distinguishes between the two fans, and between the two lamps.

\section{Results} 

\begin{center}

	\begin{tabular}{|p{0.6cm}|p{1.45cm}|p{1.55cm}|p{1.4cm}|p{1.45cm}|}

		\hline
		\bfseries{Sub} & \bfseries{SSVEP Accuracy (\%)} & \bfseries{Eye-Blink Accuracy (\%)} & \bfseries{Response Time (sec)} & \bfseries{Transfer Rate (cmd/min)}\\
		\hline
		Sub1 & 96.34 & 100 & 4.8 & 12.5\\
		Sub2 & 91.87 & 100 & 5.5 & 10.9\\
		Sub3 & 89.26 & 100 & 5.7 & 10.52\\
		Sub4 & 99.21 & 100 & 4.8 & 12.5\\	
		\hline
		\bfseries{Avg} & \bfseries{94.17} & \bfseries{100} & \bfseries{5.2} & \bfseries{11.6}\\
		\hline
	
	\end{tabular}
	\vspace*{0.1cm}

\captionof{table}{\textbf{Detection accuracy of SSVEP and eye-blink artifact and response time for all subjects.}}

\end{center}

The experiments were performed on four subjects, three males (with ages 21, 21 and 35 years) and one female (aged 20 years). The subjects were all healthy with no physical or mental disabilities. Table 1 summarizes our results. Our results are better than the state-of-the-art, giving an average of 11.6 commands per minute, which is significantly higher than \cite{A13}, the previous best result.  Also, the detection accuracy for SSVEP is on par with or better than the current state-of-the-art, while detection accuracy of the eye-blink artifact is 100\% in our experiments.

\section{Conclusion}

We have successfully implemented a robust BCI for home automation with coarse bluetooth-based indoor user localization. This system detects SSVEP response to identify the device to be controlled and uses the eye-blink artifact to confirm device selection. The system is GUI-independent and uses LEDs which flicker at distinct frequencies for each device. These LEDs provide the stimuli for SSVEP detection. This system has been tested with toggling the state of table lamp and table fan and has exhibited detection accuracy on par with and transfer rate better than the current state-of-the-art.



{\small
\bibliographystyle{ieee}
\bibliography{egbib}
}

\end{document}